\documentclass[prl,twocolumn,floatfix,showpacs ]{revtex4-1}
\UseRawInputEncoding
\usepackage{amsmath}
\usepackage{amssymb}
\usepackage{graphicx}
\usepackage{dcolumn}
\usepackage{bm}
\usepackage[colorlinks=true,linkcolor=blue,anchorcolor=balack, citecolor=cyan,urlcolor=cyan]{hyperref}
\usepackage[mathlines]{lineno}
\usepackage{ulem}
\usepackage{epstopdf}
\begin{document}
\title{Hidden piezo-Hall effect}
\author{San-Dong Guo$^{1*}$, Shi-Hao Zhang$^{2*}$ and Bingwen Zhang$^{3}$}
\thanks{\textcolor[rgb]{0.00,0.07,1.00}{San-Dong Guo: sandongyuwang@163.com;\\Shi-Hao Zhang: zhangshh@hnu.edu.cn;\\Bingwen Zhang: bwzhang@mju.edu.cn}}
\affiliation{$^{1}$School of Electronic Engineering, Xi'an University of Posts and Telecommunications, Xi'an 710121, China}
\affiliation{$^{2}$School of Physics and Electronics, Hunan University, Changsha 410082, China}

\affiliation{$^{3}$Fujian Key Laboratory of Functional Marine Sensing Materials, Center for Advanced Marine Materials and Smart Sensors, College of Material and Chemical Engineering, Minjiang University, Fuzhou 350108, China}
\begin{abstract}
Extensive studies of hidden spin polarization have established a broad class of hidden phenomena in quantum materials. Here, we introduce the hidden piezo-Hall effect, in which strain induces finite anomalous Hall conductivities (AHCs) in individual sectors while the global AHC remains zero. At zero strain, neither the full system nor its sectors exhibit an anomalous Hall effect (AHE). Under uniaxial strain, however, opposite sector AHCs emerge while their sum remains canceled. We establish this effect for a $d$-wave hidden altermagnet using a tight-binding model and first-principles calculations for the experimentally synthesized material $\mathrm{RbV_2Te_2O}$. Rotating the strain direction from the $x$ to the $y$ axis reverses the signs of the sector AHCs without generating a net Hall response. The same symmetry mechanism may also give rise to hidden piezomagneto-optical, piezo-thermal Hall, piezo-Nernst, and piezo-Ettingshausen effects. Our work extends hidden physics to strain-controlled transport and identifies $\mathrm{RbV_2Te_2O}$ as a promising candidate for experimental verification.
\end{abstract}
\maketitle
\textcolor[rgb]{0.00,0.00,1.00}{\textbf{Introduction.---}}
The anomalous Hall effect (AHE) corresponds to a transverse charge transport phenomenon that occurs in magnetic materials under zero external magnetic fields\cite{q1}. Fundamentally, the emergence of the AHE requires the breaking of time-reversal symmetry and the presence of non-zero Berry curvature around the Fermi level, where spin-orbit coupling (SOC) acts as the dominant origin of Berry curvature in momentum space. Conventionally, the AHE is considered to predominantly emerge in ferromagnetic systems with non-zero net magnetization. Nevertheless, recent research progress demonstrates that compensated magnetic systems, including altermagnets, fully compensated ferrimagnets, and noncollinear antiferromagnets\cite{q2,q3,q4,q5,q6,q7,q8}, can also support the AHE in the presence of spin-orbit magnetism (SOM)\cite{q9}. Owing to its high sensitivity to electronic band topology and magnetic ordering, the AHE has become a crucial transport characterization probe. Meanwhile, it lays a solid physical foundation for the development of low-power spintronic devices, holding promising application prospects in zero-field Hall sensors and topological electronic devices\cite{q1}.

As an emerging transport effect, the piezo-Hall effect has attracted widespread research attention in recent years\cite{q10,q11,q12}. Different from previously reported conventional Hall behaviors, this unique strain-modulated Hall response has been successfully uncovered in magnetic systems\cite{q10}. In general, the AHE is absent in strain-free systems but can be effectively triggered by external strain loading. For altermagnetic materials, the microscopic physical mechanisms of the piezo-Hall effect and its inherent correlation with piezomagnetism have been systematically clarified, and a complete set of symmetry-admissible tensor components for piezo-Hall conductivity has been summarized\cite{q11}. Furthermore, a series of feasible material candidates for realizing the piezo-Hall effect has been theoretically proposed, and several of them have been successfully synthesized in experiments, providing reliable platforms for further exploring strain-tunable Hall transport properties\cite{q12}.

\begin{figure*}[t]
    \centering
    \includegraphics[width=0.95\textwidth]{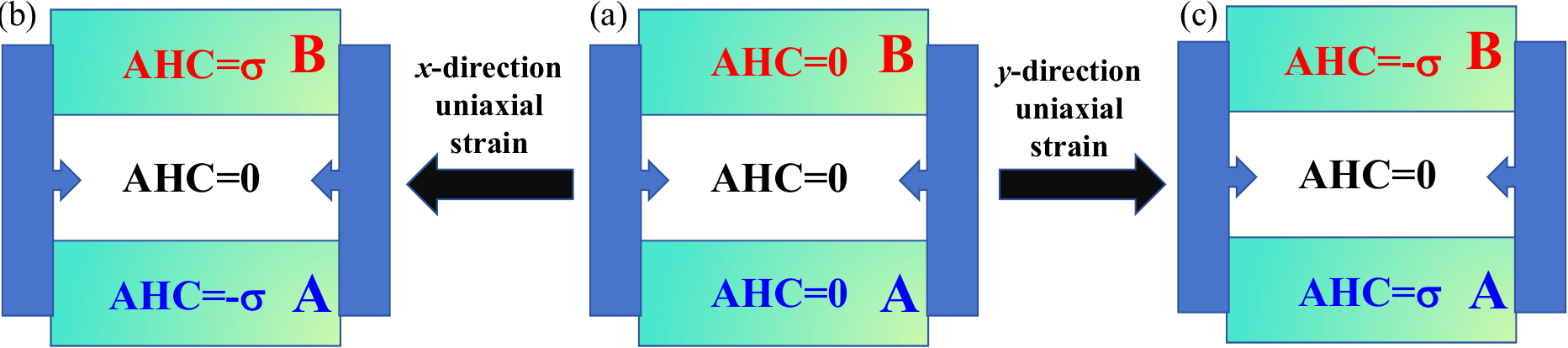}
    \caption{(Color online)  (a) in $d$-wave hidden altermagnets, the AHE is absent for sector A, sector B and the whole system under zero strain, with all corresponding AHCs equal to zero. (b)  under uniaxial $x$-direction strain, sector A and sector B exhibit AHE, but the overall system does not. Their AHCs of sector A and sector B carry opposite signs. (c)  as uniaxial strain is applied along the $y$ direction instead, the AHC signs of sector A and sector B are reversed relative to the $x$-strain configuration. The transition from panel (a) to (b) or (c) is referred to as the hidden piezo-Hall effect.}\label{a}
   \end{figure*}

Parallelly, intensive theoretical and experimental investigations on hidden spin polarization have greatly promoted the development of hidden physics\cite{h2,h3,h4,h5,h6,h7,h8,h9}. A variety of physical concepts, including hidden orbital polarization, hidden optical polarization, and hidden Berry curvature, have been successively proposed, gradually forming a distinctive research subfield of “hidden-X physics”\cite{h10,h11,h12}. However, existing studies on hidden physics mainly concentrate on the hidden characteristics of static physical quantities, while the hidden behaviors of dynamic observables remain rarely explored. This brings forward two critical scientific questions: whether dynamic physical quantities can exhibit hidden features, and whether the piezo-Hall effect can realize hidden regulation.

Against this background, we herein propose the concept of the hidden piezo-Hall effect. Specifically, no detectable AHE signal appears in either the global system or local sectors under zero-strain conditions. Upon external strain application, the global system still maintains a zero AHE response, whereas obvious AHE signals emerge in localized sectors, realizing the unique hidden modulation of the piezo-Hall effect. We analyze the electronic properties of $d$-wave hidden altermagnetism\cite{h8} and construct a tight-binding (TB) model to verify the hidden piezo-Hall effect, which is further confirmed via first-principles calculations on the experimentally synthesized $d$-wave hidden altermagnet $\text{RbV}_2\text{Te}_2\text{O}$\cite{h13,h13-1} that serves as a feasible candidate for future experimental validation.

\textcolor[rgb]{0.00,0.00,1.00}{\textbf{Hidden piezo-Hall effect.---}}
Suppose a material with layer-based degrees of freedom consists of sector A and sector B. In the absence of strain, the AHE vanishes for both the entire system and each individual sector. Under strain, the global system still exhibits no AHE, whereas AHE emerges within each sector, and their anomalous Hall conductivities (AHCs) carry opposite signs. This strain-induced local AHE is defined as the hidden piezo-Hall effect. For instance, the entire system respects  $\mathcal{PT}$ (the joint symmetry of space inversion symmetry  $\mathcal{P}$ and time-reversal symmetry  $\mathcal{T}$) symmetry regardless of whether strain is applied. Without strain, the  magnetic space groups (MSGs) of sector A and sector B forbid net magnetic moments, so the AHE is absent and the AHC vanishes. Under strain, their MSGs permit net magnetic moments, which gives rise to the AHE and a finite AHC. Sector A and sector B can correspond to   $\mathcal{PT}$-antiferromagnets and certain altermagnets without SOM\cite{q9,q9-1}, respectively.

In what follows, we take the $d$-wave hidden altermagnet as an example to explore the hidden piezo-Hall effect. This $d$-wave hidden altermagnet is primarily formed by sector A and sector B. The whole system obeys   $\mathcal{PT}$ symmetry, while the two local sectors behave as $d$-wave altermagnets\cite{h8}.
We assume that sector A and sector B possess no SOM, such that no AHE emerges and their AHCs vanish; accordingly, the global AHC of the whole system also equals zero (see \autoref{a} (a)). When uniaxial strain is applied along the $x$ direction, the whole system still retains   $\mathcal{PT}$ symmetry, yet SOM develops in sector A and sector B. The global AHE therefore remains suppressed and the global AHC stays zero, while the two local sectors exhibit AHE with AHCs of opposite signs (see \autoref{a} (b)). When the uniaxial strain is switched to the $y$ direction, the global AHC still vanishes, and the AHCs of local sector A and sector B both reverse their signs relative to the $x$-strain case (see \autoref{a} (c)). Importantly, the hidden piezo-Hall effect is not limited to hidden altermagnet systems; for example, stacking two   $\mathcal{PT}$-antiferromagnets preserves global   $\mathcal{PT}$ symmetry. Upon suitable strain application, the entire system remains  $\mathcal{PT}$-symmetric while each   $\mathcal{PT}$-antiferromagnet transforms into a fully compensated ferrimagnet, allowing the hidden piezo-Hall effect to be realized.

We next use the TB model to verify our proposal and assess its material realization in the experimentally synthesized $d$-wave hidden altermagnet $\mathrm{RbV_2Te_2O}$\cite{h13,h13-1}. The same mechanism may also occur in the related synthesized compounds $\mathrm{KV_2Se_2O}$ and $\mathrm{CsV_2Te_2O}$\cite{h14,h14-1,h14-2}. Details of the first-principles calculations are provided in the Supplemental Material (SM)\cite{SM}.

\begin{figure*}[t]
    \centering
    \includegraphics[width=0.95\textwidth]{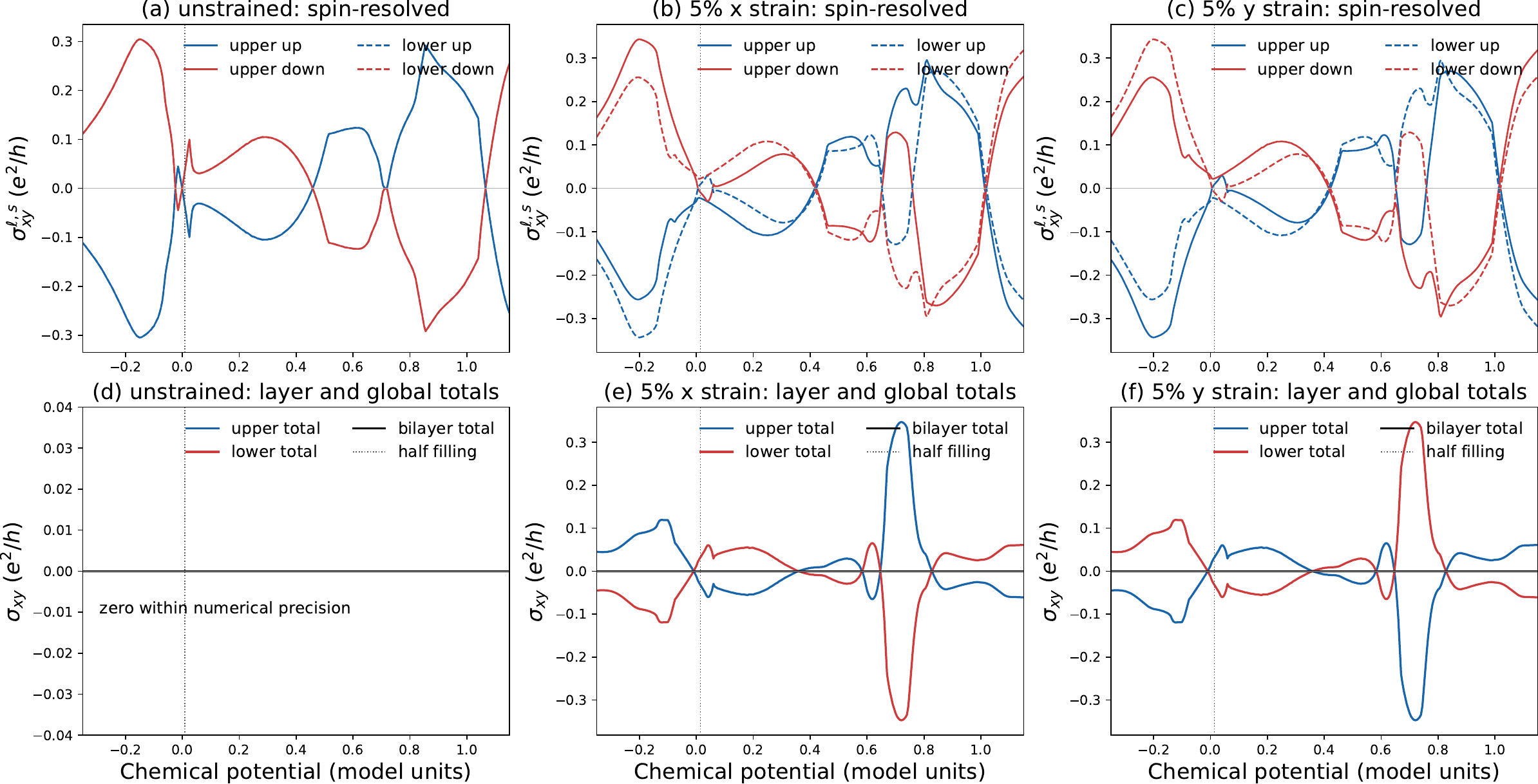}
    \caption{(Color online) Intrinsic AHC of the SOC bilayer versus chemical potential, in $e^2/h$, on a $400\times400$ momentum grid. Columns: zero strain, $5\%$ $x$ tension, and $5\%$ $y$ tension. (a)--(c) Spin-resolved lower (dashed) and  upper (solid)  layer  [sector A and sector B] responses. (d)--(f) Layer and bilayer charge responses. Vertical dotted lines mark half filling.}\label{b}
   \end{figure*}

\textcolor[rgb]{0.00,0.00,1.00}{\textbf{Model verification.---}}
To illustrate the mechanism of the hidden piezo-Hall effect, we employ a two-orbital TB model by constructing a square-lattice bilayer altermagnet ($d$-wave hidden altermagnet) with opposite exchange fields on sites $A=(0,1/2)$ and $B=(1/2,0)$\cite{q2,h8}. We retain the $d_{xz},d_{yz}$ orbitals in the basis $(A,xz;A,yz;B,xz;B,yz)$, with $\tau_i$ and $\rho_i$ acting on sublattice and orbital space and $s=\pm1$ labeling spin. The upper-layer Hamiltonian is
\begin{align}
 h_{u,s}(\bm q)={}&(\epsilon_{\bm q}+E_0)\tau_0\rho_0
 +(d_{\bm q}-s\Delta)\tau_z\rho_0\nonumber\\
 &+o_{\bm q}\tau_0\rho_z+\delta\tau_z\rho_z
 +g_{\bm q}\tau_x\rho_0\nonumber\\
 &+m_{\bm q}\tau_0\rho_x+s\frac{\lambda}{2}\tau_0\rho_y,
 \label{eq:prl_mono}
\end{align}
where $d_{\bm q}=-(t_x-t_y)(\cos q_x-\cos q_y)$ sets the $d$-wave-like anisotropy. The remaining coefficients and parameters are given in SM\cite{SM}; onsite SOC $s\lambda\rho_y/2$ preserves $s_z$ in this orbital basis. Stacking inversion-partner layers gives
\begin{equation}
 H_s(\bm q)=
 \begin{pmatrix}h_{u,s}(\bm q)&-t_\perp I_4\\
 -t_\perp I_4&h_{d,s}(\bm q)\end{pmatrix},
 \label{eq:prl_bi}
\end{equation}
where $h_{d,s}$ follows from $d_{\bm q}\to-d_{\bm q}$ and $\delta\to-\delta$. The $\lambda_i$ act in layer space; $\lambda$ without a subscript is the SOC strength. The inversion matrix $U_P=\lambda_x\tau_x\rho_0$ gives $U_PH_s^*(\bm q)U_P^\dagger=H_{-s}(\bm q)$, so $\mathcal{PT}$ pairs opposite-spin states at each $\bm q$.

The no-SOC bands reveal the symmetry sequence. In the monolayer, the valence-band spin splitting reverses sign between the $q_x$ and $q_y$ directions, producing the $d$-wave pattern of Fig.~S1. In the bilayer, $\mathcal{PT}$ restores spin degeneracy to the total bands, while the upper and lower layers retain opposite spectral spin contrasts (Fig.~S2)\cite{SM}.

For $5\%$ tension along $x$ or $y$, we scale in-plane hoppings as $t(r)=t(r_0)(r_0/r)^5$ and use a Poisson ratio of $0.25$. The $x$-strained spectrum in Fig.~S3 and $y$-strained spectrum in Fig.~S4 show that the layer spin-contrast pattern rotates with the strain axis, while the total bands remain $\mathcal{PT}$ degenerate~\cite{SM}. The broken $x\leftrightarrow y$ spin-mirror symmetry permits a layer Hall response\cite{q11}.

We compute the layer-resolved intrinsic AHC using the symmetrized current $J_a^\ell=\{P_\ell,v_a\}/2$, with $P_{u,d}=(\lambda_0\pm\lambda_z)\otimes I_4/2$ and $v_a=\partial H_s/\partial q_a$; summing the layers recovers the ordinary Berry-curvature AHC\cite{q1,SM}.
For zero temperature,
\begin{align}
 \sigma_{xy}^{\ell s}(\mu)
 &=-\frac{e^2}{h}\frac{1}{2\pi}
 \int_{\rm BZ}d^2q\sum_n f_\mu(E_{n\bm q s})
 \Omega_{n\bm q}^{\ell s},\label{eq:si_ahc}\\
 \Omega_{n\bm q}^{\ell s}
 &=-\operatorname{Im}\sum_{m\ne n}
 \frac{\langle n|J_x^\ell|m\rangle\langle m|v_y|n\rangle
 -\langle n|J_y^\ell|m\rangle\langle m|v_x|n\rangle}
 {(E_{n\bm q s}-E_{m\bm q s})^2}.
 \label{eq:si_curv}
\end{align}
 By $\mathcal{PT}$, $\sigma_{xy}^{u\uparrow}=-\sigma_{xy}^{d\downarrow}$ and $\sigma_{xy}^{u\downarrow}=-\sigma_{xy}^{d\uparrow}$, so the bilayer charge AHC vanishes.

\autoref{b} shows finite spin-resolved AHCs that cancel within each layer at zero strain. At half filling, $5\%$ $x$ tension gives $(\sigma_{xy}^{u},\sigma_{xy}^{d})=(-0.03198,+0.03198)e^2/h$; $5\%$ $y$ tension reverses the signs. The layer AHC is therefore odd under a $90^\circ$ rotation of the strain axis, while the total charge AHC remains zero. The AHCs obtained from the TB model agree with the preceding analysis, confirming the feasibility of our proposal.

\begin{figure*}[t]
    \includegraphics[width=0.95\textwidth]{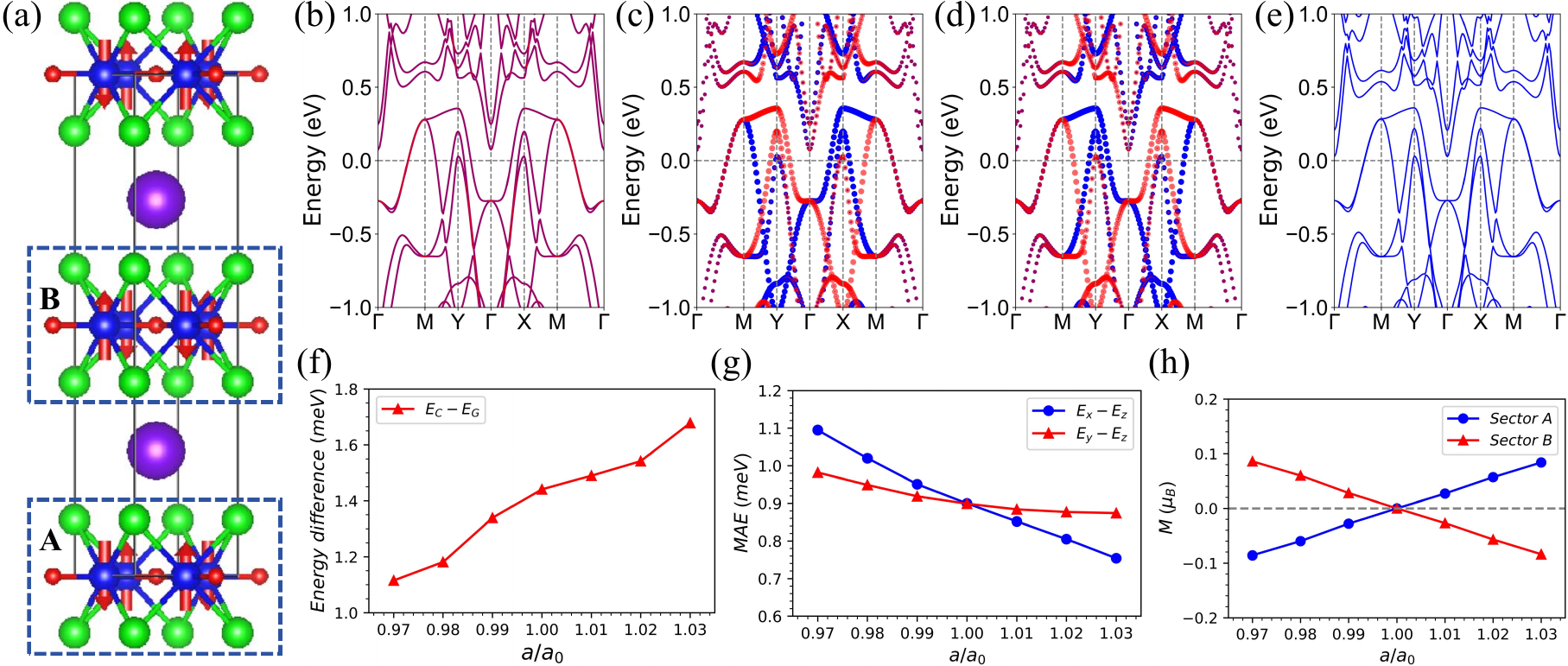}
    \caption{(Color online) For $\mathrm{RbV_2Te_2O}$: (a) crystal structure comprising two $\mathrm{V_2Te_2O}$ layers, designated sectors A and B. Purple, blue, green, and red spheres represent Rb, V, Te, and O atoms, respectively; red arrows indicate the spin directions and the G-type AFM order. Reversing the $\mathrm{N\acute{e}el}$  vector of one layer produces a C-type AFM state. (b)--(d) Spin-polarized band structure without SOC and its projections onto sectors A and B. Blue, red, and purple denote spin-up, spin-down, and spin-degenerate bands, respectively. (e) Band structure including SOC. (f) Energy difference between the C-type and G-type configurations versus $a/a_0$, with the G-type state taken as the zero-energy reference. (g) Magnetic anisotropy energy (MAE) versus $a/a_0$, with out-of-plane magnetization as the zero-energy reference. (h) Total magnetic moments of sectors A and B versus $a/a_0$.} \label{c}
\end{figure*}

\textcolor[rgb]{0.00,0.00,1.00}{\textbf{Material realization.---}}
Recent experimental studies have revealed that the quasi-two-dimensional material $\mathrm{RbV_2Te_2O}$ exhibits a G-type antiferromagnetic (AFM)  configuration and intrinsic hidden altermagnetism\cite{h13-1}. As illustrated in \autoref{c} (a), its crystal structure consists of two $\mathrm{V_2Te_2O}$ monolayers intercalated with Rb atoms, which are defined as sector A and sector B, respectively. Experimentally, each $\mathrm{V_2Te_2O}$ monolayer hosts $d$-wave altermagnetism with prominent out-of-plane magnetic anisotropy\cite{h13,h13-1}.​ Flipping the N\'eel vector of a single $\mathrm{V_2Te_2O}$ monolayer transforms $\mathrm{RbV_2Te_2O}$ into a C-type AFM state with manifest $d$-wave altermagnetic characteristics. This C-type configuration possesses an energy level comparable to the ground-state G-type configuration.​

The MSG of magnetic materials is determined by magnetization orientation, which dominates their macroscopic physical properties\cite{q9,q9-1}. For out-of-plane magnetization, $\mathrm{RbV_2Te_2O}$ adopts the   $\mathcal{PT}$-symmetric MSG $P_c4_2/mcm$, which completely suppresses the AHE. Each individual $\mathrm{V_2Te_2O}$ monolayer belongs to the MSG $P4^\prime/mm^\prime m$, which also prohibits the emergence of AHE.​
In contrast, in-plane magnetic anisotropy along the $x$ or $y$ direction changes the MSG of each monolayer to $Pm^\prime m^\prime m$, allowing the generation of AHE. Both experimental observations and theoretical calculations verify the out-of-plane magnetic anisotropy of pristine $\mathrm{RbV_2Te_2O}$\cite{h13,h13-1}, making it a reliable platform to validate our theoretical proposal.​

The spin-polarized band structures of $\mathrm{RbV_2Te_2O}$ with and without SOC, as well as the corresponding band projections on sector A and sector B, are displayed in \autoref{c} (b-e). Owing to the preserved   $\mathcal{PT}$ symmetry in G-type AFM $\mathrm{RbV_2Te_2O}$, the electronic bands remain spin-degenerate regardless of SOC effects. Projected band results demonstrate that both sectors exhibit $d$-wave altermagnetism but possess opposite spin polarization behaviors. Moreover, the metallic nature of both sectors enables strain-induced AHE measurements without carrier doping. In addition, SOC only weakly modulates the electronic states near the Fermi level.​

We apply uniaxial strain to modulate the symmetry of $\mathrm{RbV_2Te_2O}$, aiming to tailor the MSG of the global system and local sectors and realize the hidden piezo-Hall effect. The strain magnitude is defined as $a/a_0$ and $b/b_0$ for deformations along the $x$ and $y$ axes, respectively.​
Under uniaxial strain, the global lattice transforms to the MSG $P_cccm$. The G-type AFM structure with out-of-plane magnetization still maintains  $\mathcal{PT}$ symmetry, leading to zero global AHE response. Nevertheless, local sectors undergo symmetry transition to the MSG $Pm^\prime m^\prime m$, which supports finite AHE signals. This symmetry analysis solidifies the feasibility of achieving the hidden piezo-Hall effect in $\mathrm{RbV_2Te_2O}$.​

The above analysis presupposes the retention of G-type AFM order and out-of-plane magnetic anisotropy under strain. We further perform first-principles calculations to confirm these prerequisites. As presented in \autoref{c} (f, g), strained $\mathrm{RbV_2Te_2O}$ stabilizes the G-type AFM ground state and sustains out-of-plane magnetic anisotropy within the entire strain range. Global   $\mathcal{PT}$ symmetry ensures zero net magnetization under strain. However, broken $C_4$ symmetry destroys the altermagnetic order of local sectors and drives them into ferrimagnetic states\cite{gsd} with opposite  but equal-magnitude magnetic moments, as quantitatively verified in \autoref{c} (h).​ For $y$-axis uniaxial strain, the C-G AFM energy difference, magnetic anisotropy energy, and sector-resolved magnetic moments show negligible changes compared with the $x$-strain case, with only a coordinate swap of the $x$ and $y$ axes required for physical quantity conversion.

\begin{figure*}[t]
    \centering
    \includegraphics[width=0.95\textwidth]{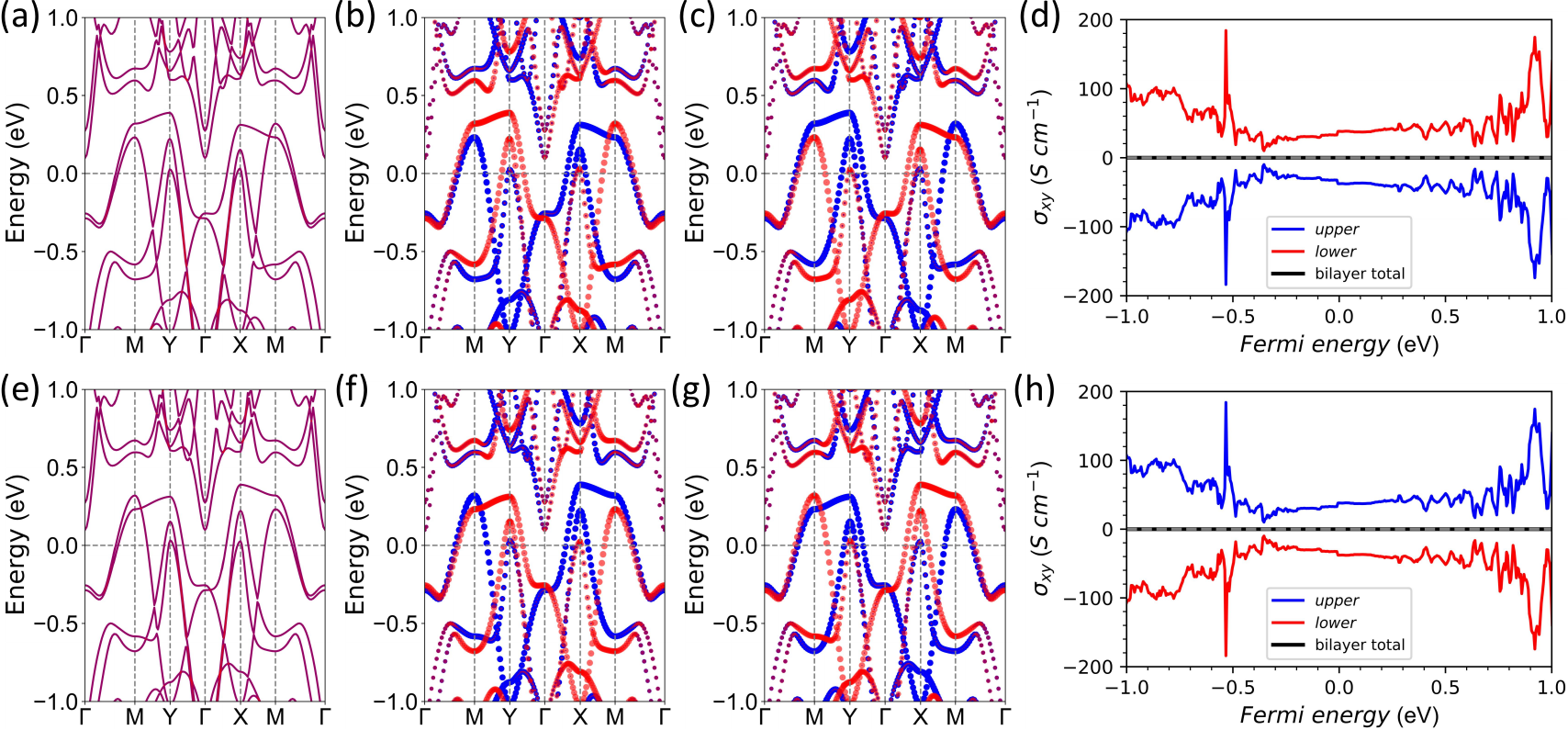}
    \caption{(Color online)  For $\mathrm{RbV_2Te_2O}$ at $a/a_0=1.01$, spin-polarized band structures calculated without SOC, together with their projections onto lower and  upper layers (sector A and sector B), along the $x$ direction [(a)--(c)] and the $y$ direction [(e)--(g)]. The sector-resolved AHCs as a function of Fermi energy (chemical potential) along the $x$ and $y$ directions is shown in (d) and (h), respectively. In [(a)--(c) and (e)--(g)], the  blue, red, and purple denote spin-up, spin-down, and spin-degenerate bands, respectively.}\label{e}
   \end{figure*}

Finally, we evaluate the strain-induced hidden AHE for $a/a_0=1.01$.  \autoref{e} shows the spin-polarized band structures calculated without SOC, their projections onto sectors A and B, and the sector-resolved AHCs as functions of the Fermi energy for strain along the $x$ and $y$ directions. Although strain preserves global $\mathcal{PT}$ symmetry and hence spin degeneracy of the total bands, the sector projections display opposite spin asymmetries, indicating equal and opposite sector magnetic moments. Rotating the strain direction from $x$ to $y$ reverses these moments. The sector AHCs likewise have equal magnitude and opposite signs, so the net AHC remains zero. Their signs also reverse upon rotating the strain direction by $90^\circ$.  For the unstrained case, the calculated sector-resolved and global AHCs are both zero and are therefore omitted from the plot. These results agree with the symmetry analysis and TB calculations, confirming the hidden piezo-Hall effect in experimentally realized $\mathrm{RbV_2Te_2O}$ and establishing this material as a promising platform for its direct observation.

\textcolor[rgb]{0.00,0.00,1.00}{\textbf{Discussion and Conclusion.---}}
The anomalous Hall and magneto-optical (Faraday and Kerr) effects share a common Berry-curvature origin and obey closely related symmetry constraints\cite{q9,q9-1}. This correspondence suggests a hidden piezomagneto-optical effect, in which strain generates opposite sector-resolved responses while the global signal remains canceled. Because magneto-optical effects are particularly accessible in semiconductors, the artificial hidden $d$-wave altermagnetic bilayer $\mathrm{Fe_2Br_2O}$, a semiconductor\cite{gsd}, provides a promising candidate.
Although piezo-Hall and piezomagnetic responses are not directly equivalent\cite{q11}, the same symmetry framework motivates a hidden piezomagnetic effect. In $\mathrm{RbV_2Te_2O}$, for example, strain leaves the total magnetic moment zero while inducing equal and opposite moments in sectors A and B (see \autoref{c} (h)). The mechanism can likewise be extended to hidden piezo-thermal Hall, piezo-Nernst, and piezo-Ettingshausen effects.

In summary, we have introduced the hidden piezo-Hall effect, in which uniaxial strain induces finite and opposite sector-resolved AHCs while the global AHC remains zero. We established this mechanism for $d$-wave hidden altermagnets using a TB model and confirmed it through first-principles calculations for synthesized $\mathrm{RbV_2Te_2O}$. The same effect should be accessible in other members of the $\mathrm{XV_2Y_2O}$ family ($X=\mathrm{K,Rb,Cs}$; $Y=\mathrm{S,Se,Te}$), including experimentally synthesized $\mathrm{KV_2Se_2O}$ and $\mathrm{CsV_2Te_2O}$, both of which exhibit hidden altermagnetism with G-type AFM order\cite{h14,h14-1,h14-2}. Hidden piezo-Hall transport therefore offers a route to manipulate local Berry-curvature responses under strain while preserving zero net magnetization, with potential applications in low-dissipation spintronics.

\begin{acknowledgments}
This work is supported by  the National Natural Science Foundation of China  (Grant No.12674079). We are grateful to Shanxi Supercomputing Center of China, and the calculations were performed on TianHe-2.
\end{acknowledgments}

\end{document}